\documentclass[twocolumn]{aa}

\usepackage{graphicx}
\usepackage{txfonts}
\usepackage{natbib}
\bibpunct{(}{)}{;}{a}{}{,} 

\def\Pdot{\dot{\rm P}}
\def\DeltaP{\left<\Delta P\right>}

\begin{document}

\titlerunning{WET Observations of BPM~37093}

\title{Whole Earth Telescope observations of BPM~37093: a seismological
test of crystallization theory in white dwarfs\thanks{Based on
observations obtained at: Observat\'orio do Pico dos Dias (OPD) Brazil,
the European Southern Observatory (ESO) Chile, South African Astronomical
Observatory (SAAO), Mt.~John University Observatory (MJUO) New Zealand,
Siding Spring Observatory (SSO) Australia, and Cerro Tololo Inter-American
Observatory (CTIO), a division of the National Optical Astronomy
Observatories, which is operated by the Association of Universities for
Research in Astronomy, Inc. under cooperative agreement with the National
Science Foundation.}}

\authorrunning{Kanaan et al.}

\author{A.~Kanaan\inst{1} \and A.~Nitta\inst{2} \and D.~E.~Winget\inst{3}
\and S.~O.~Kepler\inst{4} \and M.~H.~Montgomery\inst{5,3} \and
T.~S.~Metcalfe\inst{6,3} \and H.~Oliveira\inst{1} \and L.~Fraga\inst{1}
\and A.~F.~M.~da Costa\inst{4} \and J.~E.~S.~Costa\inst{4} \and
B.~G.~Castanheira\inst{4} \and O.~Giovannini\inst{7} \and
R.~E.~Nather\inst{3} \and A.~Mukadam\inst{3} \and S.~D.~Kawaler\inst{8}
\and M.~S.~O'Brien\inst{8} \and M.~D.~Reed\inst{8,9} \and
S.~J.~Kleinman\inst{2} \and J.~L.~Provencal\inst{10} \and
T.~K.~Watson\inst{11} \and D.~Kilkenny\inst{12} \and
D.~J.~Sullivan\inst{13} \and T.~Sullivan\inst{13} \and
B.~Shobbrook\inst{14} \and X.~J.~Jiang\inst{15} \and B.~N.~Ashoka\inst{16}
\and S.~Seetha\inst{16} \and E.~Leibowitz\inst{17} \and
P.~Ibbetson\inst{17} \and H.~Mendelson\inst{17} \and E.~G.~Mei{\v
s}tas\inst{18} \and R.~Kalytis\inst{18} \and D.~Ali{\v s}auskas\inst{19}
\and D.~O'Donoghue\inst{12} \and D.~Buckley\inst{12} \and
P.~Martinez\inst{12} \and F.~van Wyk\inst{12} \and R.~Stobie\inst{12} \and
F.~Marang\inst{12} \and L.~van Zyl\inst{12} \and W.~Ogloza\inst{20} \and
J.~Krzesinski\inst{20} \and S.~Zola\inst{20,21} \and P.~Moskalik\inst{22}
\and M.~Breger\inst{23} \and A.~Stankov\inst{23} \and R.~Silvotti\inst{24}
\and A.~Piccioni\inst{25} \and G.~Vauclair\inst{26} \and N.~Dolez\inst{26}
\and M.~Chevreton\inst{27} \and J.~Deetjen\inst{28} \and
S.~Dreizler\inst{28,29} \and S.~Schuh\inst{28,29} \and
J.~M.~Gonzalez~Perez\inst{30} \and R.~{\O}stensen\inst{31} \and
A.~Ulla\inst{32} \and M.~Manteiga\inst{32} \and O.~Suarez\inst{32} \and
M.~R.~Burleigh\inst{33} \and M.~A.~Barstow\inst{33}}

\institute{~} 

\date{Received 2004 April 21; accepted 2004 October 31}

\abstract{BPM~37093 is the only hydrogen-atmosphere white dwarf currently
known which has sufficient mass ($\sim$1.1~M$_\odot$) to theoretically
crystallize while still inside the ZZ~Ceti instability strip ($T_{\rm
eff}\sim$12,000~K). As a consequence, this star represents our first
opportunity to test crystallization theory directly. If the core is
substantially crystallized, then the inner boundary for each pulsation
mode will be located at the top of the solid core rather than at the
center of the star, affecting mainly the average period spacing. This is
distinct from the ``mode trapping'' caused by the stratified surface
layers, which modifies the pulsation periods more selectively. In this
paper we report on Whole Earth Telescope observations of BPM~37093
obtained in 1998 and 1999. Based on a simple analysis of the average
period spacing we conclude that a large fraction of the total stellar mass
is likely to be crystallized.

}

\maketitle


\footnotetext[1]{Departamento de F\'{\i}sica Universidade Federal de Santa
                 Catarina, C.P. 476, 88040-900, Florian{\'o}polis, SC, Brazil}
\footnotetext[2]{Apache Point Observatory, 2001 Apache Point Road, P.O.  
                 Box 59, Sunspot, NM 88349, USA}
\footnotetext[3]{Department of Astronomy, 1 University Station Stop C1400,
                 University of Texas, Austin, TX 78712, USA}
\footnotetext[4]{Instituto de F\'{\i}sica, Universidade Federal de Rio 
                 Grande do Sul, C.P. 10501, 91501-970 Porto Alegre, RS, Brazil}
\footnotetext[5]{Institute of Astronomy, University of Cambridge,
                 Madingley Road, Cambridge CB3 0HA, UK}
\footnotetext[6]{Harvard-Smithsonian Center for Astrophysics, USA}
\footnotetext[7]{Departamento de Fisica e Quimica, UCS, Brazil}
\footnotetext[8]{Dept. of Physics \& Astronomy, Iowa State University, USA}
\footnotetext[9]{Astronomy Dept., Southwest Missouri State University, USA}
\footnotetext[10]{Dept. of Physics \& Astronomy, University of Delaware, USA}
\footnotetext[11]{Southwestern University, Georgetown, TX, USA}
\footnotetext[12]{South African Astronomical Observatory, South Africa}
\footnotetext[13]{Victoria University of Wellington, New Zealand}
\footnotetext[14]{Chatterton Astronomy Dept., University of Sydney, Australia}
\footnotetext[15]{Astronomical Observatory, Academy of Sciences, China}
\footnotetext[16]{Indian Space Research Organization, India}
\footnotetext[17]{Wise Observatory, Tel-Aviv University, Israel}
\footnotetext[18]{Institute of Theoretical Physics \& Astronomy, Lithuania}
\footnotetext[19]{Vilnius University, Lithuania}
\footnotetext[20]{Mt.~Suhora Observatory, Cracow Pedagogical University, Poland}
\footnotetext[21]{Astronomical Observatory, Jagiellonia University, Poland}
\footnotetext[22]{Copernicus Astronomical Center, Poland}
\footnotetext[23]{Institut f\"ur Astronomie, Universit\"at Wien, Austria}
\footnotetext[24]{INAF -- Osservatorio Astronomico di Capodimonte, Italy}
\footnotetext[25]{Dipartimento di Astronomia, Universit\`a di Bologna, Italy}
\footnotetext[26]{Universit\'{e} Paul Sabatier, Observatoire
                  Midi-Pyr\'{e}n\'{e}es, France}
\footnotetext[27]{Observatoire de Paris-Meudon, France}
\footnotetext[28]{Institut f\"{u}r Astronomie und Astrophysik, Germany}
\footnotetext[29]{Universit\"atssternwarte G\"ottingen, Germany}
\footnotetext[30]{University of Troms\o, Norway}
\footnotetext[31]{Isaac Newton Group, Spain}
\footnotetext[32]{Departamento de F{\'\i}sica Aplicada, Universidad de Vigo, 
                  Spain}
\footnotetext[33]{Dept. of Physics \& Astronomy, University of Leicester, UK}


\section {Introduction}

Since 1960 most astronomers have agreed that cool white dwarfs must
eventually crystallize \citep{kir60,abr60,sal61}. The process
theoretically begins when the electrostatic interaction between the ions
becomes much larger than the thermal energy. This effect is based on such
well known physics that it has become widely accepted without ever having
been tested empirically.

BPM~37093 is a ZZ~Ceti star \citep{kan92} with an unusually high mass
\citep[1.10~$M_\odot$,][]{ber04}. White dwarfs this massive are subject to
much higher pressures and densities in their cores, and we expect a
1.0~$M_\odot$ white dwarf to begin crystallizing at temperatures within or
above the ZZ~Ceti instability strip \citep{woo92,win97}. Our goal in
observing BPM~37093 with the Whole Earth Telescope \citep[WET,][]{nat90}
was to obtain seismological data to determine whether or not the stellar
interior is crystallized. The fundamental objective was simply to detect
as many independent pulsation modes as possible, and then to compare the
observed frequencies with those calculated from white dwarf models that
have been artificially crystallized to various degrees.

For years we have faced a troubling ambiguity between the effects of the
crystallized mass fraction ($M_{\rm cr}$), the hydrogen layer mass
($M_H$), and the stellar mass ($M_*$) and effective temperature ($T_{\rm
eff}$).  Changes to these four characteristics of white dwarf models can
all modify the average spacing between the calculated pulsation periods
\citep[see][ eq.~7]{mw99}. Fortunately, the latter two quantities can be
constrained by fitting model atmospheres to spectroscopic observations,
but the others can only be determined through asteroseismology. Recent
improvements in our theoretical description of the composition transition
zones between the stratified surface layers in our models
\citep{cor02,alt03} have helped to reduce the degeneracy between $M_H$ and
$M_{\rm cr}$. However, the huge number of possible parameter combinations,
and the need for an efficient method of exploring them, remained serious
obstacles to progress until recently \citep*{mc03,mmk04}.

\begin{table}
\begin{center}
\caption{Journal of observations in 1998 (XCOV\,16).\label{tab1}}
\begin{tabular}{llccr}
\hline\hline
Run     & Telescope  & Date        &  Start    & Length \\
        &            &             &   (UT)    & (s)    \\
\hline
 an-0069 & CTIO 1.5m & 1998 Apr 17 & 02:42:20 & 23540 \\
 an-0070 & CTIO 1.5m & 1998 Apr 17 & 23:30:30 & 37460 \\
 an-0071 & CTIO 1.5m & 1998 Apr 18 & 23:16:00 & 20630 \\
 an-0072 & CTIO 1.5m & 1998 Apr 19 & 07:42:20 &  6790 \\
 an-0073 & CTIO 1.5m & 1998 Apr 20 & 23:20:10 & 37350 \\
  dj-001 & SAAO 1.9m & 1998 Apr 21 & 21:04:00 & 19400 \\
 an-0074 & CTIO 1.5m & 1998 Apr 21 & 23:25:40 & 36770 \\
  dj-002 & SAAO 1.9m & 1998 Apr 22 & 17:32:00 & 20150 \\
 an-0075 & CTIO 1.5m & 1998 Apr 22 & 23:20:40 & 36860 \\
   ro107 &  OPD 1.6m & 1998 Apr 23 & 03:57:00 &  5430 \\
  dj-003 & SAAO 1.9m & 1998 Apr 23 & 17:18:00 & 37330 \\
   ro108 &  OPD 1.6m & 1998 Apr 23 & 22:22:30 &  4700 \\
ap2498q2 & MJUO 1.0m & 1998 Apr 24 & 09:02:10 & 31110 \\
  dj-004 & SAAO 1.9m & 1998 Apr 24 & 17:29:00 & 37350 \\
 an-0076 & CTIO 1.5m & 1998 Apr 24 & 23:43:30 & 35230 \\
   ro111 &  OPD 1.6m & 1998 Apr 25 & 00:12:10 &  6650 \\
ap2598q1 & MJUO 1.0m & 1998 Apr 25 & 10:20:20 &  6060 \\
ap2598q2 & MJUO 1.0m & 1998 Apr 25 & 12:50:00 & 14280 \\
  dj-005 & SAAO 1.9m & 1998 Apr 25 & 17:34:00 & 33580 \\
   ro112 &  OPD 1.6m & 1998 Apr 26 & 01:50:30 &  7840 \\
   ro113 &  OPD 1.6m & 1998 Apr 26 & 04:53:30 &  4210 \\
ap2698q1 & MJUO 1.0m & 1998 Apr 26 & 08:06:00 & 30410 \\
  dj-006 & SAAO 1.9m & 1998 Apr 26 & 17:14:00 & 37880 \\
 an-0077 & CTIO 1.5m & 1998 Apr 27 & 03:17:20 & 21770 \\
ap2798q1 & MJUO 1.0m & 1998 Apr 27 & 13:17:30 & 37460 \\
  dj-007 & SAAO 1.9m & 1998 Apr 27 & 19:21:00 & 38030 \\
 gv-0522 &  ESO 2.2m & 1998 Apr 28 & 01:01:10 & 24570 \\
  db9801 & SAAO 1.9m & 1998 Apr 28 & 19:24:00 & 37900 \\
  db9802 & SAAO 1.9m & 1998 Apr 29 & 19:14:00 & 32880 \\
 gv-0523 &  ESO 2.2m & 1998 Apr 29 & 23:21:00 & 21770 \\
  db9803 & SAAO 1.9m & 1998 Apr 30 & 17:04:50 & 39130 \\
 gv-0524 &  ESO 2.2m & 1998 Apr 30 & 23:27:00 & 13310 \\
  db9804 & SAAO 1.9m & 1998 May 01 & 21:08:00 & 31020 \\
  db9805 & SAAO 1.9m & 1998 May 02 & 17:21:30 & 37320 \\
  db9806 & SAAO 1.9m & 1998 May 03 & 17:17:10 & 33510 \\
\hline
\end{tabular}
\end{center}
\end{table}


Like the cool DAV G~29-38 \citep{kle98}, BPM~37093 exhibits irregular
modulations in the amplitudes of its pulsation modes. On one occasion {\it
all} of the modes vanished below the detection threshold of $\sim$1~mma
\citep{kan98}. However, the modes that disappeared were observed to
reappear later with the same pulsation frequencies. This gives us
confidence that we can learn more about this star by using the full set of
frequencies that have been observed over time---a concept known as
``time-ensemble'' asteroseismology, pioneered by \cite{kle98}. In this
paper we report WET observations obtained in 1998 and 1999, and we use the
identified pulsation periods to define a range for the average period
spacing. This allows us to constrain $M_H$ and $M_{\rm cr}$ by following
the analysis of \cite{mw99} with an updated prescription for the envelope
composition transition zones.

\section{Observations}

In 1996 and 1997, observations of BPM~37093 were obtained from the 0.9~m
telescope at CTIO and the 1.6~m telescope at Observat\'orio Pico dos Dias
(OPD, Brazil) respectively. The two initial goals of these observations
were: 1) to identify as many pulsation modes as possible, to help
constrain asteroseismological model fitting, and 2) to find stable
pulsation modes suitable for measuring the rate of period change
($\Pdot$), as has been done for other white dwarfs
\citep{ckw99,kep00a,muk03}. Further analysis (see sect.~\ref{ANALYSIS})
revealed that no modes were stable enough to use for $\Pdot$ measurements.
After these two attempts to obtain single-site data on BPM~37093, it
became clear that we would be unable to resolve the pulsation spectrum of
this star from a single observatory. By 1997 we had already accumulated
more than 100 hours of photometry on BPM~37093, which led to the
identification of only 4 pulsation modes with highly variable amplitudes
\citep{kan98}.

BPM~37093 was chosen as the southern primary target for a Whole Earth
Telescope campaign (XCOV\,16) in 1998, and again in 1999 (XCOV\,17) to
coincide with simultaneous observations using the Hubble Space Telescope
(HST). A journal of observations for the data obtained for these two
campaigns is shown in Tables \ref{tab1} and \ref{tab2}. Overall, we
obtained more than 142 hours of data in April-May 1998 with a duty cycle
of 50\% during the central 10 days, and an impressive 180 hours in April
1999 with a better duty cycle of 65\% during the central 10 days. The
latter observations included almost complete coverage during the two
scheduled HST visits near the middle of the campaign, and preliminary
results were reported by \cite{nit00}.

\begin{table}
\begin{center}
\caption{Journal of observations in 1999 (XCOV\,17).\label{tab2}}
\begin{tabular}{llccr}
\hline\hline
Run       & Telescope  & Date        &  Start    & Length \\
          &            &             &   (UT)    & (s)    \\
\hline
 saccd001 & SAAO 0.75m & 1999 Apr 06 & 23:43:00 & 10680 \\
 saccd002 & SAAO 0.75m & 1999 Apr 07 & 21:16:33 &  3960 \\
 ap0899q1 &  SSO  1.0m & 1999 Apr 08 & 10:38:00 & 20130 \\
 ap0899q2 &  SSO  1.0m & 1999 Apr 08 & 16:34:00 &  3620 \\
 ap0899q3 &  SSO  1.0m & 1999 Apr 08 & 17:52:30 &  5460 \\
 ap0999q1 &  SSO  1.0m & 1999 Apr 09 & 10:15:20 & 29000 \\
 tsm-0033 & CTIO  1.5m & 1999 Apr 10 & 02:10:00 & 26100 \\
 ap1099q1 &  SSO  1.0m & 1999 Apr 10 & 09:04:40 & 36680 \\
 saccd003 & SAAO 0.75m & 1999 Apr 10 & 18:00:00 & 22060 \\
 tsm-0035 & CTIO  1.5m & 1999 Apr 11 & 05:55:00 &  6620 \\
 ap1199q1 &  SSO  1.0m & 1999 Apr 11 & 08:59:50 & 37110 \\
 saccd004 & SAAO 0.75m & 1999 Apr 11 & 18:09:00 & 33300 \\
 ap1299q1 &  SSO  1.0m & 1999 Apr 12 & 11:03:00 & 29900 \\
 saccd005 & SAAO 0.75m & 1999 Apr 12 & 18:18:30 & 32700 \\
 tsm-0042 & CTIO  1.5m & 1999 Apr 13 & 00:43:00 & 31440 \\
 ap1399q1 &  SSO  1.0m & 1999 Apr 13 & 09:35:00 &  9300 \\
 ap1399q2 &  SSO  1.0m & 1999 Apr 13 & 12:29:00 & 25000 \\
 saccd006 & SAAO 0.75m & 1999 Apr 13 & 18:17:20 & 16920 \\
 tsm-0043 & CTIO  1.5m & 1999 Apr 13 & 23:48:00 & 34410 \\
 ap1499q1 &  SSO  1.0m & 1999 Apr 14 & 09:14:00 & 36700 \\
 saccd007 & SAAO 0.75m & 1999 Apr 14 & 19:33:25 & 21840 \\
 tsm-0044 & CTIO  1.5m & 1999 Apr 14 & 23:42:00 & 34560 \\
 saccd008 & SAAO 0.75m & 1999 Apr 15 & 17:23:45 & 20250 \\
 tsm-0045 & CTIO  1.5m & 1999 Apr 16 & 00:08:30 & 32730 \\
 tsm-0046 & CTIO  1.5m & 1999 Apr 17 & 00:10:40 & 32100 \\
 ap1799q1 & MJUO  1.0m & 1999 Apr 17 & 09:34:30 &  7970 \\
 saccd009 & SAAO 0.75m & 1999 Apr 17 & 17:17:40 & 32220 \\
ita180499 &  OPD  0.6m & 1999 Apr 18 & 03:49:40 & 11000 \\
 tsm-0047 & CTIO  1.5m & 1999 Apr 18 & 04:46:00 & 15640 \\
 ap1899q1 & MJUO  1.0m & 1999 Apr 18 & 07:30:00 &  7690 \\
 ap1899q2 & MJUO  1.0m & 1999 Apr 18 & 11:36:00 &  8310 \\
 ap1899q3 & MJUO  1.0m & 1999 Apr 18 & 14:04:30 & 14400 \\
 tsm-0048 & CTIO  1.5m & 1999 Apr 19 & 00:04:30 & 32330 \\
 ap1999q1 & MJUO  1.0m & 1999 Apr 19 & 08:04:20 &  6950 \\
 ap1999q4 & MJUO  1.0m & 1999 Apr 19 & 15:05:50 & 10500 \\
ita190499 &  OPD  0.6m & 1999 Apr 19 & 22:15:10 & 32700 \\
    ro123 &  OPD  1.6m & 1999 Apr 20 & 01:23:40 & 18800 \\
ita200499 &  OPD  0.6m & 1999 Apr 20 & 03:33:24 &  8420 \\
ita210499 &  OPD  0.6m & 1999 Apr 21 & 22:15:07 & 18080 \\
\hline
\end{tabular}
\end{center}
\end{table}


The primary goal of the HST observations was to use the limb darkening
method devised by \cite{rob95} to provide an independent determination of
the spherical degree ($\ell$) for each pulsation mode. Unfortunately, this
method could be applied only to the modes with the highest amplitudes, and
the observations were not sensitive enough to distinguish between $\ell=1$
and $\ell=2$. In this paper we infer $\ell$ based only on the average
period spacing between modes.

\begin{figure}
\centering\includegraphics[width=8.8cm]{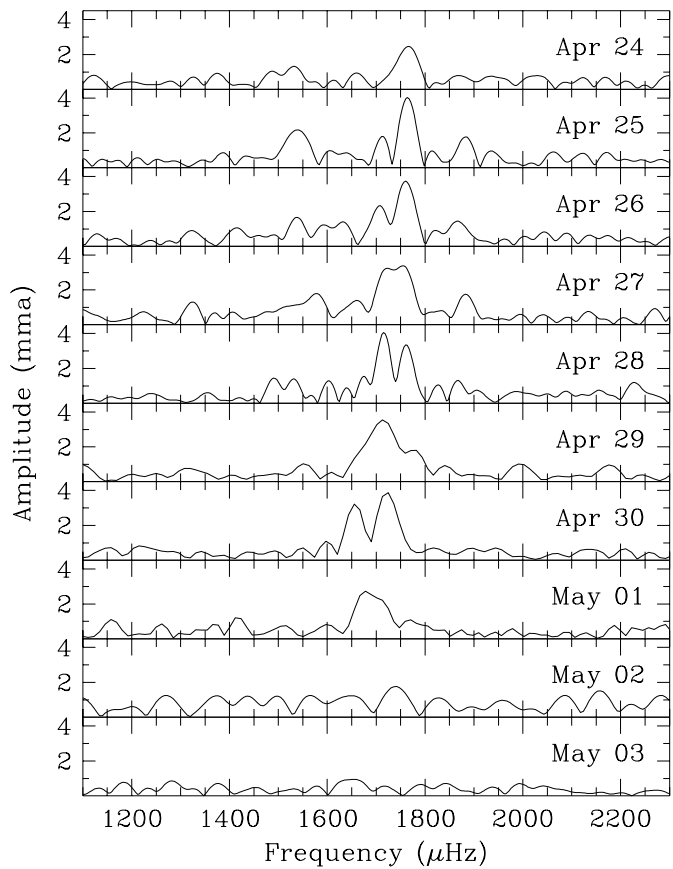}
\caption{Nightly FTs of BPM~37093 during ten nights of observations from 
CTIO in 1996. The amplitudes are highly variable on short timescales. All 
of the modes vanish below the detection threshold on the final night.}
\label{fig1}
\end{figure}

\section{Frequency Analysis \label{ANALYSIS}}

One of the initial goals of our observations was to identify pulsation
modes stable enough to measure the rate of period change ($\Pdot$). As a
white dwarf star cools over time, the slow change in the thermal structure
should lead to a detectable increase in the pulsation periods. We expect
that the periods in a crystallizing star should change more slowly than in
other ZZ~Ceti stars because the associated release of latent heat causes
the temperature to drop more slowly than in a non-crystallizing star.

\begin{figure*}
\centering\includegraphics[angle=270,width=18cm]{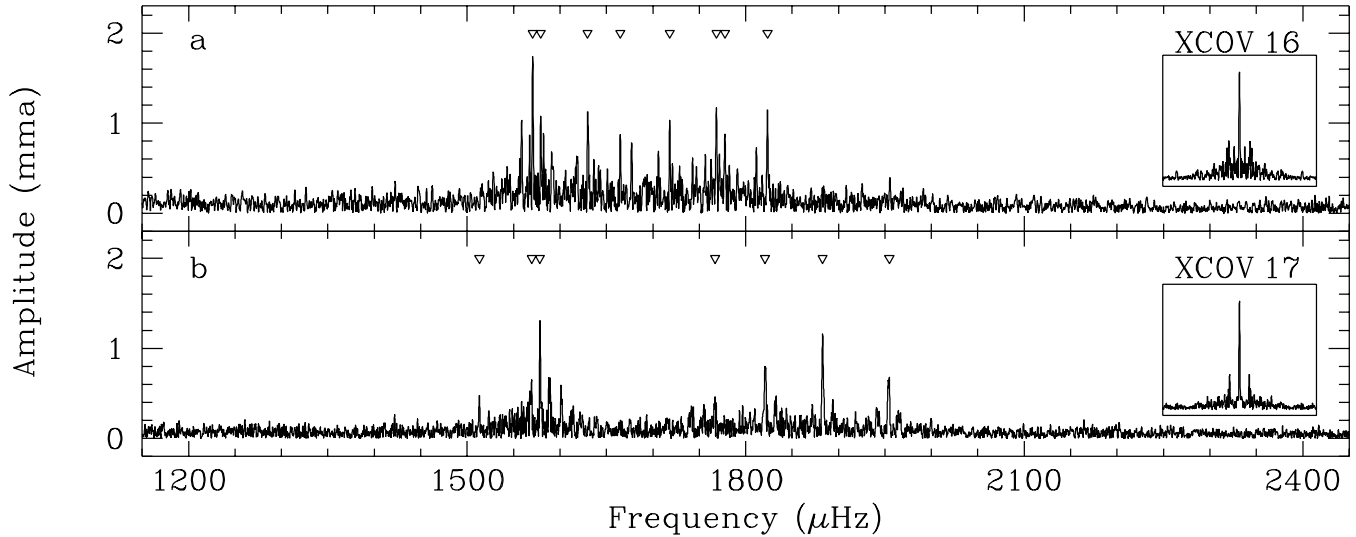}
\caption{Fourier Transforms and window functions at the same scale for the 
Whole Earth Telescope observations of the ZZ~Ceti star BPM~37093 obtained 
during {\bf a)} the XCOV\,16 campaign in 1998, and {\bf b)} the XCOV\,17 
campaign in 1999.}
\label{fig2} 
\end{figure*}

However, the observations from CTIO in 1996 made it clear that $\Pdot$
measurements would not be possible for BPM~37093. In Fig.~\ref{fig1}
we show the evolution of daily Fourier Transforms (FTs), night by night
during the 10 night run. On the tenth night, the pulsations vanished below
the detection threshold. Comparing the other panels of Fig.~\ref{fig1}, it
is also clear that the amplitudes of the detected modes are highly
variable. \cite{kan98} demonstrated that these amplitude changes must be
intrinsic to the star, rather than due to the beating of closely spaced
modes, since there was no correlation between the changes in amplitude
and phase. Such intrinsic behavior has been reported for other ZZ~Ceti 
stars on a timescale of months \citep{kle98}, but not from night to night
as observed in BPM~37093. These short-timescale variations will lead
us to derive {\it average} amplitudes from the FTs of long data sets,
but will still allow us to identify the pulsation modes which have the
highest amplitudes and the longest lifetimes.

The FTs of the two WET campaigns on BPM~37093 are shown in
Fig.~\ref{fig2}. In each panel we have marked all of the peaks that exceed
4 times the observational noise level. During XCOV\,16 we detected many
more modes than were evident in the 1996 observations---of the 8
frequencies detected, only 2 coincide with those seen previously.  
Comparing this with XCOV\,17, it seems clear that the detection of more
independent modes in XCOV\,16 can be attributed to the star itself, rather
than to differences in our detection threshold. During the second WET
campaign on BPM~37093 we detected 7 frequencies, of which 4 coincide with
others seen previously. Only the 512~s and 661~s modes are unique to the
data obtained for XCOV\,17. The full list of frequencies detected in these
two WET campaigns is shown in Table \ref{tab3}, along with the mode
identifications of \cite*{mmk04}\footnote[34]{For the 582~s mode, we adopt 
the $\ell=2$ identification} for those modes found in the preliminary 
analysis of \cite{kan00}.

\begin{table}[ht]
\begin{center}
\caption{Frequencies detected from two WET campaigns.\label{tab3}}
\begin{tabular}{rcccllr}
\hline\hline
 XCOV & Frequency & Period & $\left<\mathrm{Amplitude}\right>$ & 
                         \multicolumn{3}{c}{Identification} \\
      & ($\mu$Hz) &   (s)  &   (mma)   & $k$ & $\ell$ & $m$ \\
\hline
17~~~ & 1513.2    &  660.8 &    0.475  &     &        &     \\
\raisebox{-2pt}[0pt][0pt]{$\lceil$}
17~~~ & 1569.5    &  637.2 &    0.650  &     &        &     \\
\raisebox{2pt}[0pt][0pt]{$\lfloor$}
16~~~ & 1570.6    &  636.7 &    1.741  &  34 &   2    &$-$1 \\
\raisebox{-2pt}[0pt][0pt]{$\lceil$}
17~~~ & 1578.5    &  633.5 &    1.310  &  34 &   2    &$+$1 \\
\raisebox{2pt}[0pt][0pt]{$\lfloor$}
16~~~ & 1579.2    &  633.2 &    1.081  &     &        &     \\
16~~~ & 1629.9    &  613.5 &    1.131  &  18 &   1    &  0  \\
16~~~ & 1664.9    &  600.7 &    0.875  &  32 &   2    &  0  \\
16~~~ & 1718.2    &  582.0 &    1.032  &  31 &   2    &  0  \\
\raisebox{-2pt}[0pt][0pt]{$\lceil$}
17~~~ & 1767.1    &  565.9 &    0.458  &     &        &     \\
\raisebox{2pt}[0pt][0pt]{$\lfloor$}
16~~~ & 1768.5    &  565.5 &    1.174  &  30 &   2    &$-$1 \\
16~~~ & 1777.6    &  562.6 &    0.879  &  30 &   2    &$+$1 \\
\raisebox{-2pt}[0pt][0pt]{$\lceil$}
17~~~ & 1820.8    &  549.2 &    0.801  &     &        &     \\
\raisebox{2pt}[0pt][0pt]{$\lfloor$}
16~~~ & 1823.5    &  548.4 &    1.149  &  29 &   2    &  0  \\
17~~~ & 1882.9    &  531.1 &    1.156  &  28 &   2    &  0  \\
17~~~ & 1954.1    &  511.7 &    0.679  &  27 &   2    &  0  \\
\hline
\end{tabular}
\end{center}
\end{table}


We can understand the erratic behavior of BPM~37093 by considering
previous observations of other ZZ~Ceti pulsators. \cite{kle98} documented
very similar results for the star G~29-38, and proposed that each set of
observations can provide a subset of the full spectrum of normal modes.
The fundamental idea behind this hypothesis is that the underlying
structure of the star does not change on the timescales of the observed
amplitude variations. Instead, something in the excitation mechanism must
select certain modes and exclude others in a manner that varies over time.
Setting aside the question of what specifically causes these amplitude
modulations, we only need to assume that the observed frequencies are, in
each case, normal modes of oscillation that can be described by spherical
harmonics. Such an assumption rests on a firm body of evidence
\citep{rkn82,kep00b,cle00}, and is also supported by the fact that modes
which disappear below the detection threshold are observed to reappear
later {\it with the same frequencies}.

\section{Interpretation}

We adopt for our analysis the full set of frequencies that have been
observed in BPM~37093 over time, giving preference to the WET campaigns for
their superior frequency precision. For the isolated frequencies we assume
that each mode has an azimuthal order $m=0$ \citep[see][]{met03}, and for
the modes consisting of two closely-spaced frequencies we use the average of
the two. For modes that were observed in both campaigns, we use the
frequency from the observation with the highest amplitude. This leads to a
total of 7 independent modes that have been identified as $\ell=2$ by
\cite*{mmk04}. The average period spacing between consecutive radial 
overtones for these 7 modes is $\DeltaP=17.6\pm1.1$~s. This implies that the 
$\ell=2$ identification should be correct for the majority of the modes, 
since an $\ell=1$ identification would yield a much larger period spacing 
\citep[$\DeltaP\sim 30$~s,][]{mw99}.

\cite{mw99} performed a detailed study of the various model parameters
that could affect the average period spacing for $\ell=2$ modes. They
defined a scaling relation for $\DeltaP$ which had contributions from
variations to four parameters, including the stellar mass and effective
temperature ($M_*, T_{\rm eff}$) and the hydrogen layer mass and
crystallized mass fraction ($M_H,M_{\rm cr}$). Fortunately, $M_*$ and
$T_{\rm eff}$ can both be constrained by spectral profile fitting. The
most recent estimates for BPM~37093 come from \cite{ber04}, who found
$M_*=1.10~M_\odot$ and $T_{\rm eff}=11\,730$~K. The values of the other
parameters can only be determined through asteroseismology, but
\cite{mw99} found a troubling degeneracy between $M_H$ and $M_{\rm cr}$
that could not be broken using only the observed $\DeltaP$.

To get beyond the degeneracy, we need to use the {\it individual}
pulsation periods in addition to $\DeltaP$. Fundamentally, this is
possible because variations to $M_{\rm H}$ change the individual periods
through ``mode trapping'', while variations to $M_{\rm cr}$ affect mainly
the average period spacing. If BPM~37093 is partially crystallized, the
inner boundary for each pulsation mode will be located at the top of the
solid core rather than at the center of the star. This reduces the size of
the resonant cavity, increasing the average period spacing and modifying
the periods of {\it all} of the modes. By contrast, the hydrogen layer
produces a sharp chemical gradient somewhere near the surface. Any
individual mode whose eigenfunction is large in this region can interact
with the chemical gradient and become ``trapped''---its period will be
shifted more than the periods of other modes. Clearly, each mode will be
modified in a different manner through this interaction.

\begin{figure}
\centering\includegraphics[width=8.8cm]{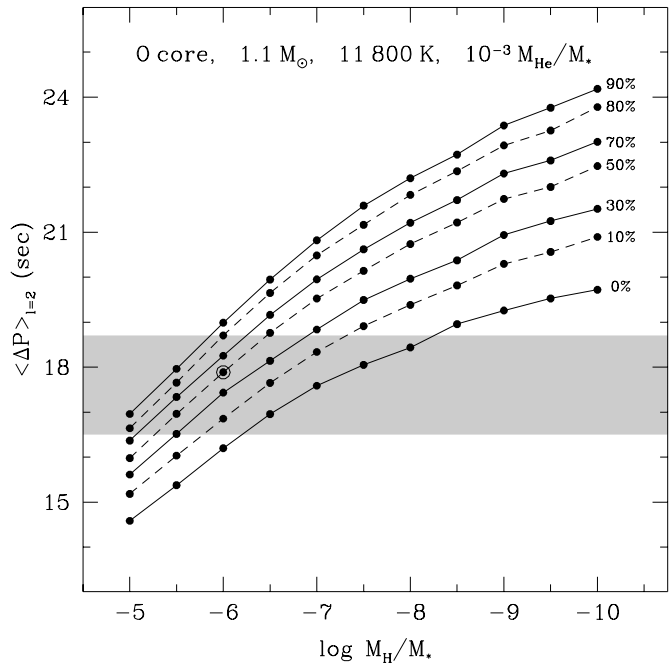}
\caption{The average period spacing of a small grid of models with various 
combinations of $M_H$ and $M_{\rm cr}$. The 1$\sigma$ range of the 
observed average period spacing for BPM~37093 is shown as a shaded area,
and the circled point indicates the model with the smallest rms difference
between the observed and calculated periods (see text for details).}
\label{fig3} 
\end{figure}

\cite{mw99} understood that the signatures of $M_H$ and $M_{\rm cr}$ are
in principle distinct. However, by focusing only on the average period
spacing they were unable to place strong constraints on either $M_H$ or
$M_{\rm cr}$. Our present models are somewhat improved: \citeauthor{mw99}
used hydrogen profiles that were derived assuming diffusive equilibrium in
the trace element approximation. This produced unrealistically sharp
chemical gradients at the base of the hydrogen layer, leading to stronger
mode trapping in their models. This was demonstrated by \cite{cor02}, who
compared models that assumed diffusive equilibrium in the trace element
approximation with models that computed the abundance profiles based on
time-dependent diffusion calculations. In a recent extension of this work
to massive ZZ~Ceti stars, \cite{alt03} described an improved method of
calculating diffusive equilibrium profiles that compare favorably with the
fully time-dependent results (see their Fig.~18). We have incorporated
this method of computing the hydrogen abundance profiles into the code
used by \cite{mw99}. However, since the sharpness of the hydrogen
transition zone should mainly affect the mode trapping properties of the
models, we expect that our new average period spacings will differ only
slightly from those computed by \cite{mw99}.

As a simple illustration of the potential of our observations, we
calculated $\DeltaP$ for a small grid with various combinations of $M_H$
and $M_{\rm cr}$. We fixed the mass, temperature, and helium layer
thickness to the values used for Fig.~10b of \cite{mw99}, but we assumed a
uniform O core. We show this grid of models in Fig.~\ref{fig3} with the
shaded 1$\sigma$ range of the average period spacing from the WET
observations of BPM~37093. As expected, the average period spacing of the
0\% crystallized model is virtually identical to that found by
\cite{mw99}. However, due to the different assumed C/O profiles, the
crystallized curves have shifted with respect to the results of
\cite{mw99}.

Unfortunately, the degeneracy between $M_H$ and $M_{\rm cr}$ is still
present, but we have not yet used the hidden third dimension of the grid:
at each point we have also computed the root-mean-square differences
between the observed and calculated periods ($\sigma_{\rm P}$), to exploit
the information contained in the {\it individual} modes. Although the
observed range of $\DeltaP$ can accommodate values of $\log(M_H/M_*)$
between $-5$ ($M_{\rm cr}=80$-90\%) and $-8$ (uncrystallized), not all
hydrogen layer masses are equally successful at reproducing the individual
modes. The model with $\log(M_H/M_*)=-6$ and $M_{\rm cr}=50$\% has
$\sigma_{\rm P}=1.08$~s, which is substantially better than anything else
in this small grid (the next best model has $\sigma_{\rm P}=1.70$~s).
A theoretical model with this same set of structural parameters is
expected to be between 66-92\% crystallized for a C/O mixture, or 
even more crystallized if the core is composed of an O/Ne mixture 
\citep[see][ for some recent calculations]{cor04}.

Of course, the individual modes also contain information about the mass,
temperature, helium layer, and core composition. If we have fixed these
parameters incorrectly, it is likely that we have found a {\it locally
optimal} match to the observed periods rather than the global solution.  
\cite{mw99} recognized this difficulty, and discussed the need for an
automated procedure to search this enormous parameter-space. Such a
procedure, based on a parallel genetic algorithm, has recently been
developed \citep{mc03} and applied to this problem. The initial results
from this large-scale exploration of the models have been published
separately by \cite*{mmk04}, who present a more detailed discussion of
the initial model fitting results.

\section{Discussion}

The Whole Earth Telescope has once again deciphered the complex pulsation
spectrum of an astrophysically interesting white dwarf star. BPM~37093 is
the only known ZZ Ceti star massive enough to allow a seismological test
of crystallization theory, and previous attempts to understand it from
single site observations were not successful. The superior frequency
precision and extended coverage of two WET campaigns finally allowed us to
document a series of 9 independent pulsation modes in this star. While it
is always useful to search for additional frequencies to help constrain
the model fitting, the observational requirements of this project have
now largely been satisfied.

The limiting factor in our ability to test crystallization theory through
asteroseismology of BPM~37093 is now computational. The initial study of
\cite*{mmk04} was limited to several fixed masses and core compositions,
but {\it all} of their optimal models suggested that a large fraction of
the core is crystallized---a result that is qualitatively supported by our
simple analysis of the average period spacing. We have only fit $M_{\rm
cr}$ to the nearest 0.1 $M_*$, but \cite{mw99} showed that the pulsations
are sensitive to changes of 0.01 $M_*$ in this parameter. As computers get
faster, we should be able to extend the genetic algorithm based model
fitting method to treat $M_{\rm cr}$, $M_*$ and the composition of the
liquid portion of the core as fully adjustable parameters. In the
meantime, we should apply the same fitting procedure to lower mass ZZ~Ceti
stars that are not expected to be crystallized, as a check of the method.

The Sloan Digital Sky Survey (SDSS) has recently discovered several new
ZZ~Ceti stars that may also be massive enough to test crystallization
theory \citep{muk04}. All of the new pulsators in the SDSS sample are
significantly fainter than the previously known white dwarfs, so larger
telescopes will probably be required to resolve their pulsation spectra.
Several current and planned space missions (MOST, \citealt{wal03}; COROT,
\citealt{bag02}) promise to revolutionize the field of asteroseismology in
the coming decade. However, none of them is likely to contribute to our
understanding of pulsating white dwarfs stars, which are too faint to be
observed by these satellites. The Whole Earth Telescope remains the
instrument of choice for these fainter targets. The availability of larger
ground-based telescopes may allow the WET concept to be extended to 2~m
and 4~m class instruments. With a growing harvest of new objects from
SDSS, the future looks bright for white dwarf asteroseismology.

\begin{acknowledgements}

This work was partially supported by NASA, NSF, Brazilian institutions
FAPERGS, CNPq, CAPES, and FINEP, and by the Smithsonian Institution
through a CfA Postdoctoral Fellowship. PM is partially supported by Polish
KBN grants 5 P03D 012 20 and 5 P03D 030 20.

\end{acknowledgements}

\end{document}